\documentclass[letterpaper,12pt,final,3p,times,longbibliography]{elsarticle}
\usepackage{graphicx,epsfig,url,lineno}
\usepackage{amssymb,fontenc,times,mathptmx}
\usepackage{wrapfig,rotating}
\usepackage[usenames]{color}
\usepackage{subcaption}
\usepackage{siunitx}
\usepackage[colorlinks = true,
            linkcolor = black,
            urlcolor  = blue,
            citecolor = blue,
            anchorcolor = blue]{hyperref}
\usepackage{tikz}
\usepackage{amsfonts}
\usepackage{systeme}
\usepackage{amsmath}   % for cases construct
\usepackage{enumitem}
\usepackage{datatool-base}

\journal{LCWS 2023 Proceedings}

\def \mZ {m_{\mathrm{Z}}}

\def \ee {\mathrm{e}^{+}\mathrm{e}^{-}}
\def \electron {\mathrm{e}^{-}}
\def \positron {\mathrm{e}^{+}}

\def \mumu {\mu^+ \mu^-}

\def \eemmg { \mathrm{e}^{+}\mathrm{e}^{-} \rightarrow \mu^+ \mu^- (\gamma)}

\def \invfb {\mathrm{fb}^{-1}}
\def \invab {\mathrm{ab}^{-1}}

\def \sqrtsp {\sqrt{s}_{p}}

\def \g1z {g_{1}^{\mathrm{Z}} }

\def \Ebm {E_{\mathrm{b}}^{-}}
\def \Ebp {E_{\mathrm{b}}^{+}}

\def \Ebn {E_{\mathrm{b}}^{\text{nom}}}

\def \poneTbf { \mathrm{\hat{\mathbf{p}}}^{\text{T}}_{1} }
\def \ptwoTbf { \mathrm{\hat{\mathbf{p}}}^{\text{T}}_{2} }

\def \p12mag {|\vec{p}_{12}|}
\def \xp {x^{\prime}}
\def \Ep {E^{\prime}}
\def \fpeak { f_{\mathrm{peak}} }

\begin{document}

\begin{frontmatter}
\title{{\Large{\bf Further investigation of dilepton-based center-of-mass energy measurements 
at e$^{+}$e$^{-}$ colliders}}}

\author{Graham W. Wilson}
\address{Department of Physics and Astronomy, University of Kansas, \\
  Lawrence, KS 66045, USA}

%spell out LaTeX symbols to ease arXiv posting.
\begin{abstract}
Methods for measuring the absolute center-of-mass energy using dileptons from 
e$^{+}$e$^{-}$ 
collision events
are further developed with an emphasis on accelerator, detector, and physics limitations.
We discuss two main estimators, the lepton momentum-based center-of-mass 
energy estimator, 
$\sqrt{s}_{p}$, 
discussed previously, 
and new estimators for the electron and positron colliding beam energies, 
denoted 
$E^{\text{C}}_{-}$ 
and 
$E^{\text{C}}_{+}$. 
In this work we focus on the
underlying limitations from beam energy spread, detector resolution, and the modeling 
of higher-order QED radiative corrections associated 
with photon emissions originating from 
initial-state-radiation (ISR), final-state-radiation (FSR), and their interference. 
We study the consequent implications for the potential of these methods
at center-of-mass energies ranging from 90 GeV to 1 TeV relevant to a number of potential
accelerator realizations in the context of measurements of masses of 
the Z, W, H, top quark, and new particles. The statistical importance 
of the Bhabha channel for Higgs factories is noted.
Additional extensive work on improving the modeling of 
the luminosity spectrum including the use of copulas is also reported.
\end{abstract}

\end{frontmatter}

\begin{center}
{\it Submitted to the Proceedings of the International Workshop on \\ Future Linear Colliders (LCWS 2023) }
\end{center}

\section{Introduction}
Recent work on measuring the absolute center-of-mass energy using dileptons (specifically dimuons) from collision events 
at future $\ee$ colliders was reported in~\cite{BMadison}. For ILC beam conditions at $\sqrt{s}=250$~GeV 
and using full simulation of the ILD detector concept, it was shown that
a statistical precision of 2.1~ppm on the center-of-mass energy scale could be achieved with a 
2.0~$\invab$ unpolarized data-set with the dimuon channel alone using the $\sqrtsp$ center-of-mass energy 
estimator based on the measurement of the muon momenta. In this work, we report on a number of new 
developments that advance our understanding of the utility and potential of this method.
\section{Colliding Beam Energies using $\ee \to \mu_{1}^{+} \mu_{2}^{-} (\gamma)$}
One can infer knowledge of 
the $\electron$ and $\positron$ beam energies of the actual collision 
from the muons alone under the assumption of one collinear but {\it undetected} ISR photon. 
The energy and $z$-direction{\footnote{The $z$-axis is defined as the axis that bisects the outgoing electron beam-line axis and the negative of the 
incoming positron beam-line axis.} longitudinal momentum ($E, p_{z}$) conservation equations in the lab frame are
\begin{flalign}
E_{-} + E_{+} &= E_{1} + E_{2} + |p_{\gamma}^{z}| / \cos(\alpha/2) \: \: ,\\
(E_{-} - E_{+}) \cos(\alpha/2) &= p_{1}^{z} + p_{2}^{z} + p_{\gamma}^{z} \: \: ,
\end{flalign}
where $E_{-}$ and $E_{+}$ are the energies of the colliding beam particles in the 
lab frame, with crossing-angle, $\alpha$, 
($E_{1}, p_{1}^{z}$) and ($E_{2}, p_{2}^{z}$) are the 
energies and $z$-direction longitudinal momenta of the anti-muon and muon respectively, and $p_{\gamma}^{z}$, is the $z$-component of the 
momentum of the collinear (with one of the beams) undetected ISR photon. 

These can be solved for $E_{-}$ and $E_{+}$, leading to:
\begin{flalign}
E_{-}  &= \frac{1}{2} \left[ (E_{1} + E_{2}) +  \frac{(p_{1}^{z} + p_{2}^{z})}{\cos{(\alpha/2)}} \right] 
+ \frac{ (|p_{\gamma}^{z}| + p_{\gamma}^{z}) } {2 \cos(\alpha/2) } \: \: , \\
E_{+}  &= \frac{1}{2} \left[ (E_{1} + E_{2}) -  \frac{(p_{1}^{z} + p_{2}^{z})}{\cos{(\alpha/2)}} \right] 
+ \frac{ (|p_{\gamma}^{z}| - p_{\gamma}^{z}) } {2 \cos(\alpha/2) }  \: \: ,
\end{flalign}
where of course one has assumed the unknowable knowledge of the actual ISR photon momentum. 
Instead, dropping the ISR photon terms in each equation leads to the following estimators,
\begin{flalign}
E^{\text{C}}_{-}  &= \frac{1}{2} \left[ (E_{1} + E_{2}) +  \frac{(p_{1}^{z} + p_{2}^{z})}{\cos{(\alpha/2)}} \right] \: \: , \\
E^{\text{C}}_{+}  &= \frac{1}{2} \left[ (E_{1} + E_{2}) -  \frac{(p_{1}^{z} + p_{2}^{z})}{\cos{(\alpha/2)}} \right] \: \: ,
\end{flalign}
which rely simply on the muon measurements and neglect the undetected ISR photon.
Obviously, the errors on these estimates, $\Delta E^{\text{C}}_{\pm}$,  that are caused by dropping the ISR photon terms are: 
\begin{flalign}
\Delta E^{\text{C}}_{-}  &= E^{\text{C}}_{-} - E_{-} = 
- \frac{ (|p_{\gamma}^{z}| + p_{\gamma}^{z}) } {2 \cos(\alpha/2) } \: \: , \\
\Delta E^{\text{C}}_{+}  &= E^{\text{C}}_{+} - E_{+} =  
- \frac{ (|p_{\gamma}^{z}| - p_{\gamma}^{z}) } {2 \cos(\alpha/2) }  \: \: .
\end{flalign}

The consequence of the ISR photon induced error on $E^{\text{C}}_{\pm}$ depends on which beam emitted the photon 
as characterized by 
the sign of the ISR photon's longitudinal momentum:
\begin{flalign}
( \Delta E^{\text{C}}_{-}, \Delta E^{\text{C}}_{+} ) &= \begin{cases}
                                             (0, \: -E_{\gamma}) & p_{\gamma}^{z} < 0 \\
                                             (- E_{\gamma}, \: 0 ) & p_{\gamma}^{z} > 0 \: \: .
                                          \end{cases}
\end{flalign}
When the ISR photon is emitted from the positron beam ($p_{\gamma}^{z} < 0$), $E^{\text{C}}_{-}$ is {\it exact} under the assumption of one collinear undetected ISR photon, but $E^{\text{C}}_{+}$ can be 
very wrong (underestimated by the photon energy, $E_{\gamma} = |p_{\gamma}^{z}| / \cos{(\alpha/2)}$). 
Conversely, when the ISR photon is emitted from the electron beam ($p_{\gamma}^{z} > 0$), $E^{\text{C}}_{+}$ is {\it exact}, but $E^{\text{C}}_{-}$ can be very wrong.
Thus for each event this method allows 
exact reconstruction of one of the colliding beam particle 
energies under the one collinear ISR photon assumption.
Identifying, with certainty, which one is exact on an event-by-event basis 
is not feasible, but the method does provide 
statistical information related to the distribution of the 
colliding beam particle energies. Furthermore, we should emphasize that the obtained 
estimate, which corresponds to the actual colliding beam energy in the lab frame, 
is sensitive to the combined effect of the colliding particle's energy deviation from nominal 
arising from both the beam energy spread fluctuation and the often present beamstrahlung-induced energy loss.

Figure~\ref{fig:cbeamenergy} illustrates these distributions at 
the WHIZARD~\cite{WHIZARD,OMEGA} generator level for ILC at $\sqrt{s}=250$~GeV 
with super-imposed fits corresponding to the convolution with a single Gaussian 
of an admixture of a beta distribution and a delta function 
as described in \ref{app:GPCBT}.
The fitted Gaussian resolution parameter, $\sigma$, is very similar to the intrinsic expectation from 
beam energy spread alone of 0.190\% ($\electron$) and 0.152\% ($\positron$) indicating as expected that the 
peak regions of these distributions retain significant information relevant to characterizing the 
initial beam parameters and distributions.
\begin{figure}[!htbp]
\centering
\includegraphics[height=0.45\textheight]{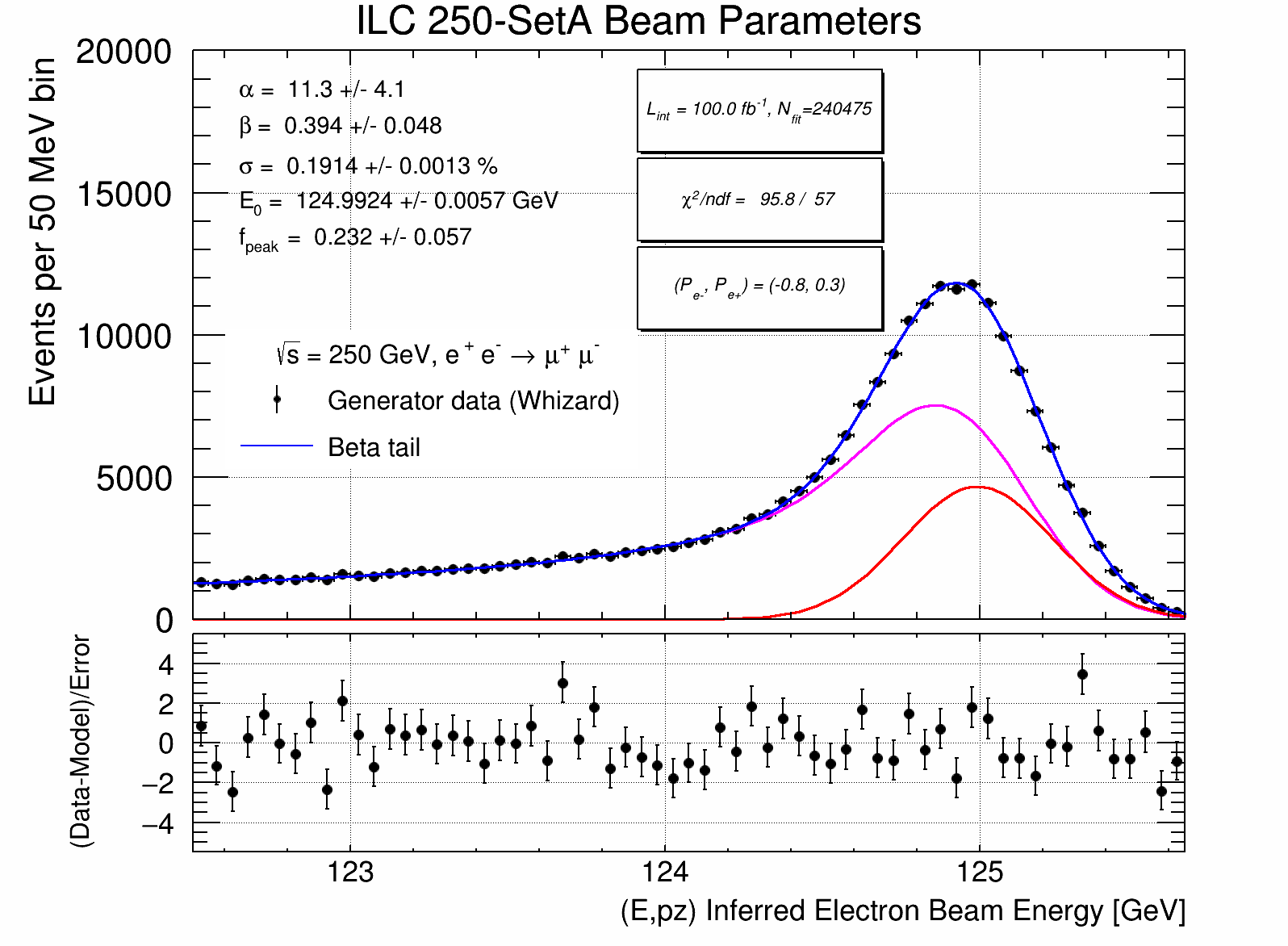}
\includegraphics[height=0.45\textheight]{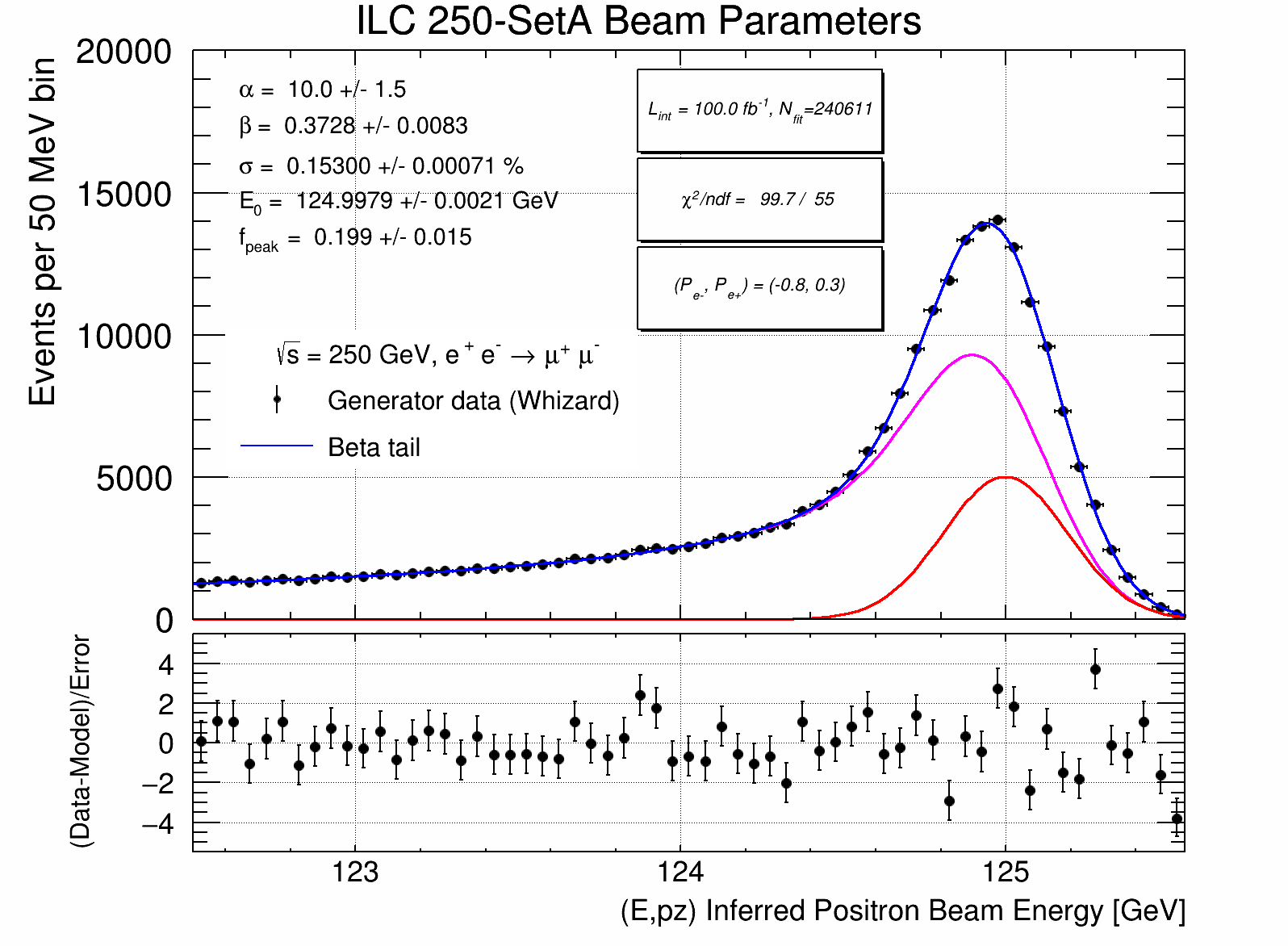}
\caption[]{\small \sl 
Distribution of the inferred colliding electron beam energy, $E^{\mathrm{C}}_{-}$ (top), and the 
inferred colliding positron beam energy, $E^{\mathrm{C}}_{+}$ (bottom),
at generator level for $\eemmg$ events with super-imposed 5-parameter fits. 
The simulation includes the effects of beam energy spread, beamstrahlung, and radiative corrections 
as included in the WHIZARD-based iLCSoft samples.
}
\label{fig:cbeamenergy}
\end{figure}
%
% Should I try to treat contamination?
%
% I should include the two plots on page 19.
\begin{table}[hbt]
\begin{center}
\begin{tabular}{|c|r|r|r|r|}
\hline 
     Observable                 &   $\sqrt{s}$ &    $\Ebm$ &    $\Ebp$  &  $2 \sqrt{ \hat{E}_{\mathrm{b}}^{-} \hat{E}_{\mathrm{b}}^{+}} \cos(\alpha/2)$\\  \hline
     WHIZARD generator level    &   3.8     &     3.2      &    2.6   &  2.1 \\  
     Generator level estimator  &   5.8     &   6.4   &  5.3   &   4.1  \\
     Detector level estimator   &   7.8     &   12.5  &  12.2  &  8.7  \\ \hline
\end{tabular}
\end{center}
\caption{Statistical precision estimates on the absolute scale of the
center-of-mass energy, $\sqrt{s}$, and the electron and positron beam energies ($\Ebm$ and $\Ebp$) 
in parts per million (ppm) for 100~$\invfb$ of $\eemmg$ events with $P(\electron)=-0.8$ and $P(\positron)=+0.3$ 
at $\sqrt{s}=250$~GeV for ILC. The statistical energy scale uncertainties are the result of fits to 
the relevant distributions with the shape parameters fixed to their best-fit values, 
but the one scale-related location parameter floating.
The $\sqrtsp$ estimator is used for the $\sqrt{s}$ estimate, and the ($E, p_{z}$)-inferred colliding 
beam energy estimators are used for $\Ebm$ and $\Ebp$. The WHIZARD generator level values represent 
fits to the true distributions of $\sqrt{s}$, $\Ebm$, and $\Ebp$ differing only from 
the pure luminosity spectrum equivalent by the convolution with the cross-section.
The generator level estimator includes effects from ISR and FSR that invalidate the assumptions.
Detector level estimates use ILD full simulation with dimuons classed 
in the gold, silver, and bronze $\sqrtsp$ resolution categories and includes the effects of acceptance, efficiency, and resolution. 
The last column gives the propagated uncertainty on the 
center-of-mass energy scale using the single beam energy scale estimates 
neglecting potential correlations.
}
\label{tab-precisions}
\end{table}

Table~\ref{tab-precisions} shows a comparison of the statistical precisions for $\sqrt{s}$ and the individual 
beam energies using 
the $\sqrtsp$, $E^{\mathrm{C}}_{-}$, and $E^{\mathrm{C}}_{+}$ 
observables at generator and detector level 
and a corresponding estimate of the center-of-mass energy from 
the individual beam energy estimates.
There are two main intrinsic effects affecting 
the WHIZARD generator level estimates for $\sqrt{s}, \Ebm, \Ebp$: the different fractional standard deviations associated 
with pure beam energy spread (0.12\%, 0.19\%, 0.15\%) 
and the different probabilities, and therefore purity 
for negligible beamstrahlung induced energy loss. 
Using the same characterization as 
in Section~\ref{sec:lumimodel}, 25.20\% of luminosity spectrum 
events are in the joint peak region, 
while 48.01\% are in the peak regions of each beam.
The statistical precision on $\sqrt{s}$ using $\sqrtsp$ is degraded 
first by physics effects (the assumptions of the $\sqrtsp$ method), 
and then additionally by detector resolution. 
A similar picture is at play for 
the $E^{\mathrm{C}}_{-}$ and $E^{\mathrm{C}}_{+}$ observables but 
there is a larger degradation than for $\sqrt{s}$. 
Some of this is the effective factor of two loss in beam particle 
numbers associated with the hemisphere ambiguity, but 
the generator estimator to detector estimator degradation 
is surprisingly big, and deserves more study.
%A more sophisticated treatment of the hemisphere ambiguity 
%issue may be able to recuperate some of the degradation, or it could be 
%simply a reflection of the 
%\input{Precisions}
\section{Beam Energy Spread}
The beam energy spread is a fundamental limitation on how well one 
can measure the center-of-mass energy. It affects 
the ``luminosity spectrum'' and induces a variable longitudinal boost. 
The fractional energy spread, for the energies of each beam and of the center-of-mass assuming Gaussian uncorrelated beams, is illustrated in Fig.~\ref{fig:BES} 
for ILC\footnote{The gray curve with no undulator is not 
the current ILC baseline; this 
could be an electron-driven positron source that avoids the undulator energy spread degradation, but would probably not allow for polarized positrons.} and FCC-ee\footnote{Updated to layout PA31-3.0 of June 2023.}.
As one can see the fractional energy spread decreases with center-of-mass energy for linear colliders and increases with 
center-of-mass energy for circular colliders.
Note beamstrahlung effects are not included in the figure 
for ILC, but they are included for FCC-ee where 
the multi-orbit effect results in effectively an 
increased pre-collision energy spread.
The ILC is described 
in~\cite{ILCInternationalDevelopmentTeam:2022izu} with an accelerator 
configuration for the Z discussed originally in detail in~\cite{Yokoya:2019rhx}; 
the same considerations are used for the $\sqrt{s}=161$~GeV parameters.
%{\bf FIXME Update FCC-ee curve}.
\begin{figure}[!htbp]
\centering
\includegraphics[height=0.45\textheight]{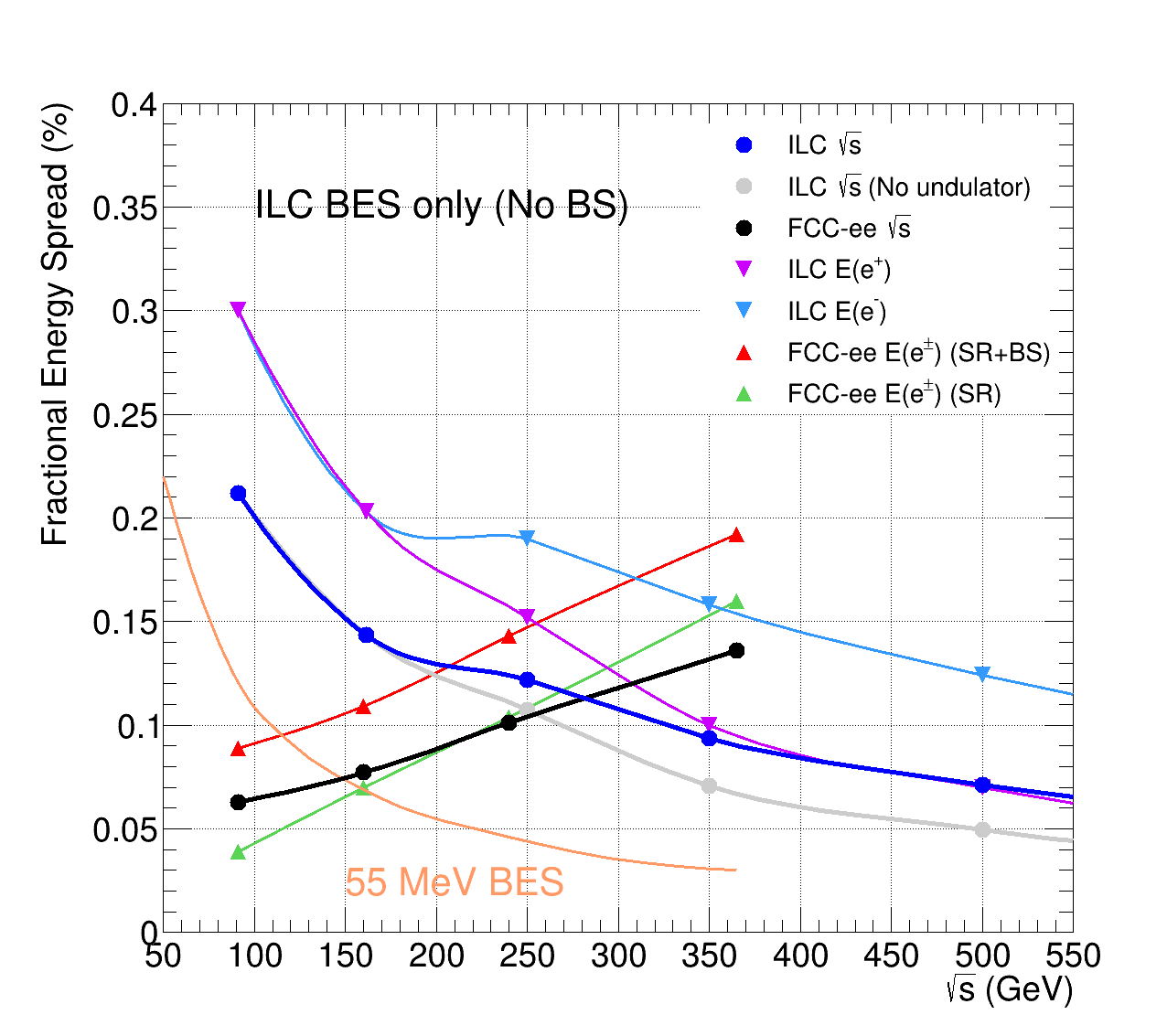}
\caption[]{\small \sl 
Fractional energy spread on the center-of-mass energy 
for ILC (dark blue), FCC-ee (black) and ILC with no undulator (gray). 
Also shown are the spreads for the individual beam energies. 
For ILC the electron energy spread (light blue) is 
degraded additionally compared to the positron (violet) 
by passage through the undulator used for positron production. 
For FCC-ee the total beam energy spread is shown in red while the synchrotron radiation component is shown in green. The orange curve with a beam energy spread of 55~MeV 
delimits the region to the lower left where resonant depolarization based 
beam energy measurement may be feasible.
}
\label{fig:BES}
\end{figure}
One sees that the center-of-mass energy spread varies 
from 0.21\% at $\sqrt{s}=91$~GeV to 0.07\% at $\sqrt{s}=500$~GeV 
for ILC, while it varies from 0.06\% at $\sqrt{s}=91$~GeV to 0.14\% at $\sqrt{s}=365$~GeV for FCC-ee, and is 0.10--0.12\% in the $\sqrt{s}=250$~GeV region for both. 
This sets corresponding targets for momentum resolution.
\section{Improved/More Versatile Modeling of the Luminosity Spectrum}
\label{sec:lumimodel}
In our previous work~\cite{BMadison} 
based on the standard ILC generator files 
we noted some deficiencies in the 
smoothness of the simulated beam-related distributions (see also Fig.~\ref{fig:cbeamenergy}).
This likely arises from the combined behavior of the specific implementation 
of the Guinea-PIG based beam 
simulation~\cite{Schulte:1997nga} modeling of the luminosity spectrum and 
the numerical sampling from this luminosity spectrum using the CIRCE2 interface in the WHIZARD event generator. 
For an introduction and in-depth discussion of luminosity spectrum issues see~\cite{Poss:2013oea}.
We identified two 
specific modeling limitations that play at least a partial role.
%in this mis-modeling. 
Firstly, the input beam energy distributions were truncated at $\pm3.0$ standard deviations, and secondly, 
the input beam energy distributions were only sampled from a limited set of 
the {\it same} 80,000 beam particles (separately for $\electron$ and $\positron$) 
in each Guinea-PIG run.

The CIRCE2 implementation uses
a purely numerical grid-based model for interfacing 
simulated luminosity spectra with event generators. However 
we have been encouraged by the relative simplicity of the ILC
luminosity spectra at $\sqrt{s}=250$~GeV and have pursued 
a parametric approach to luminosity spectrum modeling similar 
to CIRCE1~\cite{Ohl:1996fi}. In the past the observation that 
the electron and positron energy distributions are not independent 
led to arguably a premature abandonment of the parametric approach, 
which has the potential to be more versatile, accurate, and 
reproducible. To address these issues in a manner that promised to be expedient 
for the dilepton studies, and is likely of wider utility, 
we have made progress in a number of directions:

\begin{enumerate}
\item Simplifying the use of Guinea-PIG for beam simulations targeted at physics and detector 
studies by avoiding the use of input beam files and deferring the simulation of beam energy spread to a second step. In this way 
the Guinea-PIG output consists only of 
the luminosity spectrum with beamstrahlung effects. 
To simplify the simulations 
the {\tt force\_symmetric} setting is turned 
on (assumes symmetric charge distributions).
It is also straightforward to implement 
different (Gaussian) beam energy spread settings.
However it is not clear how to include explicit E-$z$ beam correlations, where $z$ is the co-moving 
longitudinal coordinate of the bunch.

This setup has been used in the development of the 
GP2X framework reported in~\cite{BMadison_LCWS2023} where 
events from Guinea-PIG runs are used for 
luminosity spectrum modeling and are mixed with events from physics event generators.

\item Parametrization of these pure beamstrahlung 
luminosity spectra using 
a model with separate two-component mixtures of beta distributions 
for the ``Body'' region and the ``Arms'' regions together with 
a delta function ``Peak'' region component. This is inspired 
by Andr\'e Sailer's ``CoPa'' parametrization~\cite{Sailer:2009zz}. 
The regions in the scaled 2-d energy distribution of 
the electron and positron, ($x_{-}, x_{+}$), 
where $x_{-} = \Ebm/\Ebn$ and $x_{+} = \Ebm/\Ebn$,
are defined by whether the fractional 
energy loss of each beam exceeds 2~ppm as illustrated in Fig.~\ref{fig:armbody}.
\item Use of copulas to parametrize the dependence issues in the pure beamstrahlung 
luminosity spectra for the Body events where both 
beams experience significant energy loss from beamstrahlung.
\item Developing a simple stochastic model for generating luminosity spectrum events including 
dependence effects (ie. electron/positron energy correlations) and beam energy spread.
\end{enumerate}

For item (2), we have used 2M Guinea-PIG simulated events 
for the ILC $\sqrt{s}=250$~GeV configuration with no beam energy spread 
using the Guinea-PIG++ implementation~\cite{Rimbault}.
We used 20 independent runs of 100,000 events 
with 500,000 macroparticles in each run to better 
avoid the noted non-stochastic issues. These 
simulated events are also used in the GP2X study.

With the region definitions 
of Fig~\ref{fig:armbody}, 25.20\% of events lie in 
the Peak region, 45.62\% in the Arms regions, 
and the remaining 29.18\% in the Body region.
We have fitted the Guinea-PIG distributions 
of $x_{\pm}$ for the events in the Body region (two entries per Body event) 
and the $x_{\pm}$ distribution using $x_{-}$ for the Arm$-$ region 
and $x_{+}$ for the Arm$+$ region (one entry per Arm event). 
The fits are done as a function of 
the transformed 
variable, $t \equiv (1 - x)^{1/\eta}$, with $\eta=4$.
The resulting fits with a double beta tail model 
described in~\ref{app:DBT} are summarized in Table~\ref{tab:lumifit}.
They are fairly reasonable\footnote{There are indications 
from the fit $\chi^{2}$ that a better description for 
the Body events may be warranted.}. 
%given the significant number of events 
%fitted. 
\begin{figure}[!htbp]
\begin{center}
\begin{tikzpicture}
%\draw[help lines, color=gray!30, dashed] (0,0) grid (5.1,5.1
\draw[help lines, color=gray!30, dashed] (0,0);
\draw[->,ultra thick] (0.0,0.0)--(5.0,0.0) node[right]{$x_{-}$};
\draw[->,ultra thick] (0.0,0.0)--(0.0,5.0) node[above]{$x_{+}$};
\draw[black] (4.5,4.0) node[left,font=\small]{Peak};
\draw[black] (4.0,2.5) node[left, rotate=90, font=\small]{Arm$+$};
\draw[black] (2.5,4.0) node[left,font=\small]{Arm$-$};
\draw[black] (2.35,1.7) node[left,font=\small]{Body};
\draw[-, thick] (3.2,0)--(3.2,5.0) node[above]{};
\draw[-, thick] (0,3.2)--(5.0,3.2) node[above]{};
\end{tikzpicture}
\end{center}
\caption[]{\small \sl 
The regions in the scaled energies of the electron 
and positron beam, $x_{-}$ and $x_{+}$, respectively. 
For the standard tests, the regions were defined using  
$1 - x_{\pm} = 2 \times 10^{-6}$, which is three orders 
of magnitude smaller than the the typical beam energy spread of 0.17\%.
}
\label{fig:armbody}
\end{figure}
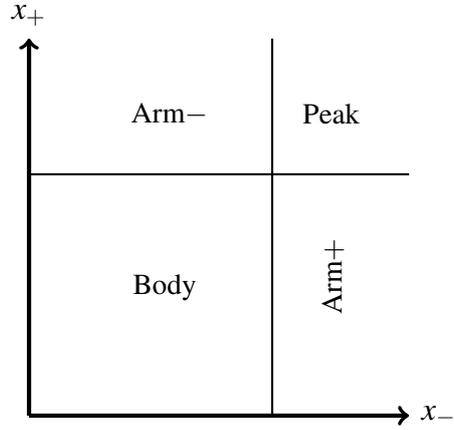
\begin{table}[!htbp]
\begin{center}
\begin{tabular}{|r|r|r|}
\hline
Parameter & Body&  Arms\\ \hline
%$\alpha_{1} - 1$  &  $12.61 \pm 0.12$ &  $14.26 \pm 0.21$  \\
%$\beta_{1} -1 $  &  $-0.6421 \pm 0.0017$ &  $-0.623 \pm 0.016$  \\ 
%$\alpha_{2} - 1$  &  $ 38.9 \pm 7.4 $ &  $30.5 \pm 1.6 $  \\
%$\beta_{2} - 1$   &  $ -0.389 \pm 0.066$ &  $-0.672 \pm 0.016$  \\ 
$\alpha_{1} $  &  $13.61 \pm 0.12$ &  $15.26 \pm 0.21$  \\
$\beta_{1}  $  &  $0.3579 \pm 0.0017$ &  $0.377 \pm 0.016$  \\ 
$\alpha_{2} $  &  $ 39.9 \pm 7.4 $ &  $31.5 \pm 1.6 $  \\
$\beta_{2} $   &  $ 0.611 \pm 0.066$ &  $0.328 \pm 0.016$  \\ 
$f_{1}$   &  $0.939 \pm 0.017$  &   $0.674 \pm 0.037$   \\  \hline
$\chi^2/\nu$ &   123.0/68       &     79.6/68    \\ \hline
$p_{\text{region}}$(\%) & $29.179 \pm 0.032$ & $45.619 \pm 0.035$ \\ \hline  
\end{tabular}
\end{center}
\caption{
Parameter values for the 5-parameter double 
beta function fits to the Body and Arms regions for ILC at $\sqrt{s}=250$~GeV. Also listed are the multinomial observed probabilities (no fit) for falling 
in each region.
% See /home/graham/gpFits
}
\label{tab:lumifit}
\end{table}
For the 29.2\% of events that lie in 
the Body region we also need to take care of 
modeling the dependence\footnote{Here we use dependence rather 
than correlation since the dependence can be more 
general than the usual linear correlation.} 
structure in the Body region's ($x_{-}, x_{+}$) bivariate distribution. 
Some dependence is expected because of the finite bunch length 
as interactions occurring from head-head collisions tend to 
have less energy loss compared with those from tail-tail 
collisions which tend to have more energy loss.

Instead of attempting to fit 
the bivariate probability distribution, $p(x_{-}, x_{+})$, in its entirety, 
we will use the copula approach to factor 
the modeling problem into two more manageable pieces. 
The first piece is the modeling of the marginal distributions and the second 
is the determination of the dependence structure using copulas. 
The marginal distributions were already found above. Thus, 
we already have a parametrization from the performed fit 
for the $x_{-}$ and $x_{+}$ marginal distributions, 
$p(x_{-})$ and $p(x_{+})$, (assumed to be equal here) where,
\begin{equation}
p(x_{-}) = \int p(x_{-}, x_{+}) dx_{+} = \int p(x_{-}, x_{+}) dx_{-} = p(x_{+}) .
\end{equation}
We do not explicitly know $p(x_{-}, x_{+})$ yet, and we will 
not need to directly find it.

Next we tackle the issue of dependence modeling using copulas. 
The techniques are applied frequently in quantitative finance but 
are far less common in the physical sciences. 
The monographs of Nelsen~\cite{Nelsen} and Joe~\cite{Joe} and references 
therein are comprehensive guides to the rich literature.  
Before delving too much into copulas let us first establish 
that there is some 
appreciable dependence to model in the problem at hand. 
There are a variety of 
statistics that can be applied to the observed $(x_{-}, x_{+})$ distribution. We have computed Blomqvist's beta (medial correlation coefficient), 
Spearman's rho (rank correlation coefficient), 
and Kendall's tau (concordance based rank correlation coefficient) 
for Guinea-PIG runs with ILC accelerator conditions 
at 91, 161, and 250 GeV (again neglecting beam energy spread). 
These statistics should all be consistent with zero if there 
is no dependence, i.e. if the $x_{-}$ and $x_{+}$ variables are independent.
Results are reported in Table~\ref{tab:stats} (see~\cite{Joe} for definitions) 
establishing clearly that for all center-of-mass energies, but 
especially for $\sqrt{s}=250$~GeV, that there is a significant, but 
relatively small, dependence present.
\begin{table}[!htbp]
\begin{center}
\begin{tabular}{|r|r|r|r|r|}
\hline
$\sqrt{s}$ (GeV) & $N$ &  $\beta_{B}$ (\%)  & $\rho_{S}$ (\%) &  $\tau_{K}$ (\%)\\ \hline
91   &  185,589 &  $1.21 \pm 0.23$  &  $1.59 \pm 0.23$ &  $1.06 \pm 0.15$  \\
161  &  347,021 &  $1.70 \pm 0.17$  &  $2.65 \pm 0.17$ &  $1.77 \pm 0.11$ \\ 
250  &  583,584 &  $3.66 \pm 0.13$  &  $5.41 \pm 0.13$ &  $3.61 \pm 0.09$  \\ \hline  
\end{tabular}
\end{center}
\caption{
Measured values of Blomqvist's beta, Spearman's rho, and Kendall's tau 
dependence statistics in the Body region for 
Guinea-PIG ILC samples with 2M events at 91, 161, 250 GeV.
The number of events in the Body region with the $1-x = 2\times 10^{-6}$ separation 
definition is given in the second column.
% See /home/graham/gpFits
}
\label{tab:stats}
\end{table}

So what is a copula? 
The copula of a random variable vector ($X_{-}, X_{+}$) is defined as 
the joint cumulative distribution function of the random variable 
vector ($U_{-}, U_{+}$),
\[ C(u_{-}, u_{+}) = \text{Pr} [U_{-} \le u_{-}, U_{+} \le u_{+}] \: \: ,\] 
where ($U_{-}, U_{+}$) is related to ($X_{-}, X_{+}$) via 
the cumulative distribution functions ($F_{-}$ and $F_{+}$)  
of the original univariate marginal distributions by 
\[ (U_{-}, U_{+})  =  (F_{-} (X_{-}), F_{+} (X_{+})) \: \: ,\]
and as usual, 
\[ F_{-} (x) = \text{Pr} [X_{-} \le x] \: \: \text{and} \: \: F_{+} (x) = \text{Pr} [X_{+} \le x]   \: \: .  \]
With this construction, 
the $C(u_{-}, u_{+})$ copula's univariate marginal distributions 
are both uniformly distributed on [0,1]. 
The copula provides the link between the marginal distributions of $x_{-}$ 
and $x_{+}$ and the joint distribution of $(x_{-}, x_{+})$.
Now we seek a copula model that fits adequately 
the dependence of the observed 
$(x_{-}, x_{+})$ distribution. After some trial and error, we chose a 3-parameter mixture model of the 1-parameter 
Clayton and Ali-Mikhail-Haq (AMH) copulas,
\[  C(u_{-}, u_{+}; \phi, \theta_{C},\theta_{AMH})  =  \cos^{2}{\phi} \: C_{1} ( u_{-}, u_{+}; \theta_{C})  + \sin^{2}{\phi} \: C_{2} (u_{-}, u_{+}; \theta_{AMH} ) \: \: . \] 
Here the $\phi$ parameter is used to adjust the weight of each copula while respecting normalization and positivity requirements, 
the Clayton copula, defined for $0 \le \theta_{C} < \infty $, is 
\[ C_{1} ( u_{-}, u_{+}; \theta_{C}) = ( u_{-}^{-\theta_{C}} + u_{+}^{-\theta_{C}} -1 )^{-1/\theta_{C}} \: \: , \]
and the AMH copula, defined for $-1 \le \theta_{AMH} \le 1$, is
\[ C_{2} ( u_{-}, u_{+}; \theta_{AMH}) = \frac{ u_{-} u_{+}} {1 - \theta_{AMH} (1 - u_{-}) (1 - u_{+} ) } \: \: .\] 

We fitted this model to 
the 250~GeV Guinea-PIG ILC sample. 
The fit was done using a pseudo-likelihood fit 
to the simulated data cast as an empirical copula 
by maximizing,
\[  {\cal{L}} ( \phi, \theta_{C},\theta_{AMH}    )  = \prod_{i=1}^{N} c(U^{i}_{-}, U^{i}_{+}; \phi, \theta_{C},\theta_{AMH}) \: \: .\]
Here $c$ is the corresponding 
probability density function obtained from differentiating the copula model. It is 
evaluated at points $(U^{i}_{-}, U^{i}_{+})$ for 
each event, $i$, computed on the unit square 
using the ranks\footnote{We resolved occasional nominal ties associated 
with the limited precision used in Guinea-PIG 
for storing post-beamstrahlung energies by ranking 
based on higher precision values where a small 
random Gaussian smearing of 0.001 ppm had been added.}, $R^{-}_{i}$ and $R^{+}_{i}$
from 1 to $N$, of the 
observed $x_{-}$ and $x_{+}$ values, 
by defining 
$U^{i}_{-} = (R^{-}_{i} - 0.5)/N$ and 
$U^{i}_{+} = (R^{+}_{i} - 0.5)/N$.
A rank of 1 corresponds to the lowest observed value for 
the corresponding $x_{\pm}$ variable.
% See ~/gpDigest/WorkFlow.txt to check
This leads to the sampled marginal distributions of 
the empirical copula being perfectly uniform 
(with no fluctuations) rather than 
just being distributed uniformly as required by the definition.
The use of ranks makes the 
determination of the copula parameters 
independent of the modeling of the original univariate marginal distributions.
The fit results with  
the obtained parameter values are 
shown in Table~\ref{tab:copulafit} and the data to model 
comparison is shown in Figure~\ref{fig:copcomp} using 
dependence deviations based 
on the functions denoted left and right tail 
concentration functions in~\cite{Venter}, which are related to the probability content in 
the lower-left and upper-right square regions of the 2-d copula distribution, and 
defined as 
\[ L(z) = \text{Pr} ( U_{-} < z, U_{+} < z)/z \: \: ,\]
and 
\[ R(z) = \text{Pr} ( U_{-} > z, U_{+} > z)/(1-z) \: \: .\]
If there is no dependence, one expects $L(z) = z$ and $R(z) = 1 - z$, 
and so the plotted variable is the deviation formed 
by subtracting the expected values under the independence hypothesis.
One sees that the left tail (region of large energy loss in both beams) 
has a significant positive dependence and also 
the right tail (small energy loss in both beams) has 
a smaller positive dependence.
The fitted model agrees well with the observed 
shape of the measured deviations from 
independence as a function of $z$. 
\begin{table}[!htbp]
\begin{center}
\begin{tabular}{|r|r|r|r|}
\hline
$\sqrt{s}$ (GeV) & $\phi$ (rad) &  $\theta_{C}$  & $\theta_{AMH}$ \\ \hline
250   &   $0.444 \pm 0.070$  &  $0.0497 \pm 0.0051$ & $0.36 \pm 0.10$  \\ \hline
\end{tabular}
\end{center}
\caption{
Values for the copula parameters fitted 
to the Guinea-PIG ILC sample with 2M events 
at 250 GeV.
% See /home/graham/gpFits
}
\label{tab:copulafit}
\end{table}
\begin{figure}[!htbp]
\centering
\includegraphics[height=0.40\textheight]{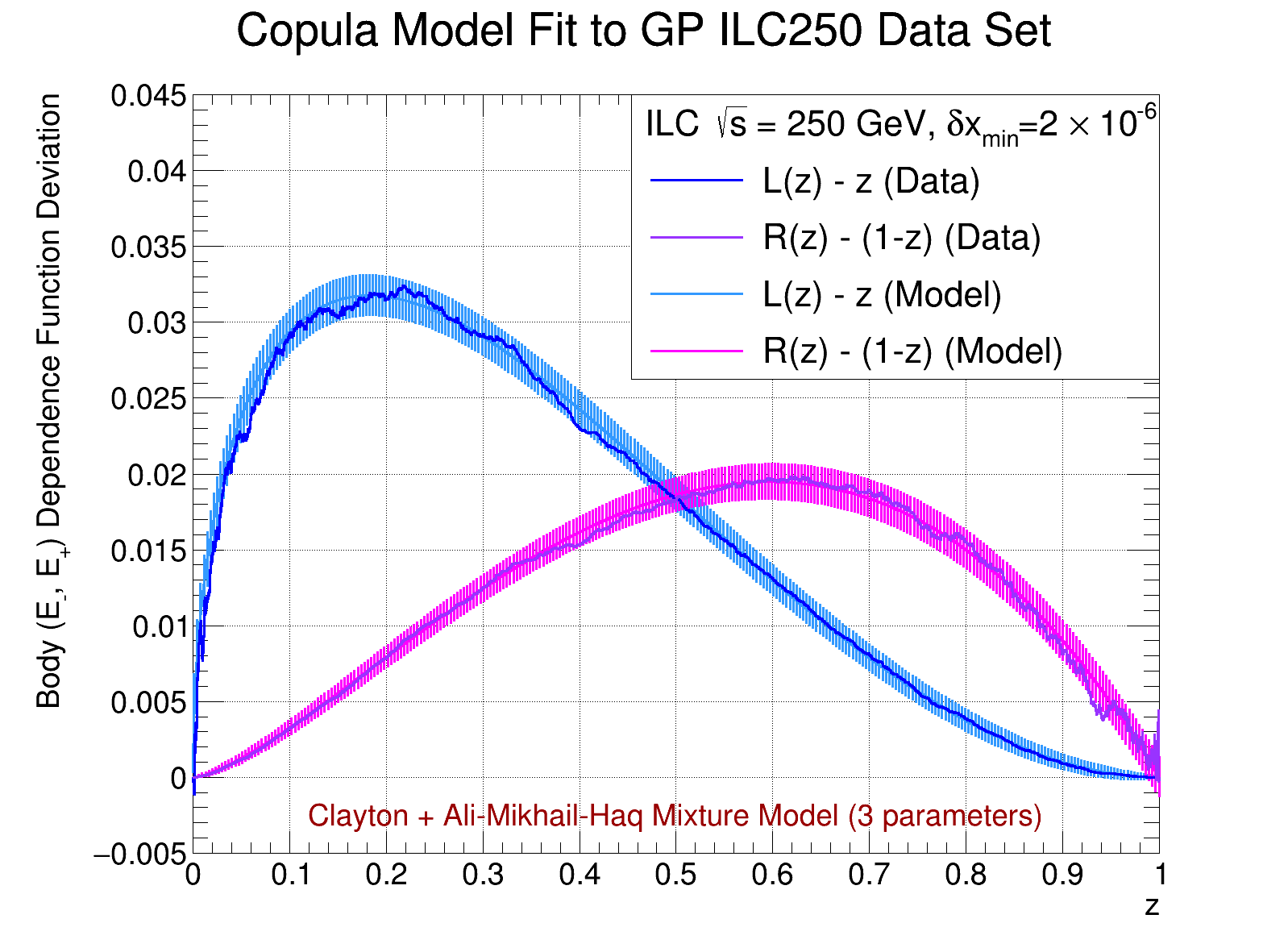}
\caption[]{\small \sl 
Comparison of the simulated data from Guinea-PIG for ILC at $\sqrt{s}=250$~GeV 
in the Body region to the predictions of the fitted model in terms of 
the dependence function deviations, $L(z) - z$ and $R(z) - (1-z)$, 
that should be zero for all $z$ in [0,1] for no dependence. 
%The $L(z), R(z)$ functions are defined in the text. 
The displayed uncertainties show the 
statistical uncertainties on the model prediction for 
the sample size of the simulated data ($N=583,584)$. 
Note given the use of cumulative distributions the bin-to-bin uncertainties are not independent.
}
\label{fig:copzplot}
\end{figure}
A goodness-of-fit statistic was calculated using chi-squared with 10,000 cells 
each with dimension of $0.01 \times 0.01$ on the unit square, 
and found $\chi^{2}/\nu = 9488.4/9797$ where the calculated number 
of degrees of freedom, $\nu$, takes into account the three fit parameters 
and the 200 constraints implicit in the empirical copula construction.
We also followed this up
%\footnote{Initial tests indicated $\chi^2$ $p$-values that were 
%surprisingly close to 1.0 and so it was decided  } 
with more 
formal testing using a parametric bootstrap based on 
the Cram\'er--von Mises statistic ($S_{n}$) following 
the methodology of Appendix A in~\cite{Genest-GoF} and 
found a $p$-value of about 98\% confirming the initial impression that 
the copula fit model fits well. As part of this bootstrap procedure we 
developed a stochastic implementation to generate events from the copula model.

We have also developed a stochastic implementation that generates events from the 
($E_{-}, E_{+}$) luminosity spectrum after beamstrahlung and beam energy spread. This 
uses the 10 parameters of the marginal distribution fits of Table~\ref{tab:lumifit}, the 
two region probability parameters, two parameters for the Gaussian beam energy 
spread of each beam, two parameters 
for potential (small) changes in energy scale of each beam and 
the three copula parameters for Body events for a total of 17 parameters.
The region (Peak, Arm$-$, Arm$+$, Body) is chosen by throwing 
a single uniform random number. 
Values of ($x_{-}, x_{+}$) are generated according to 
the defined beamstrahlung-only model including the dependence structure for Body events 
and where beams with $1-x < 2 \times 10^{-6}$ are set to $x=1.0$ (Peak and Arms events).
The ($E_{-}, E_{+}$) distribution including Gaussian beam energy spread effects is then 
obtained by smearing and potentially shifting the ($x_{-}, x_{+}$) values and scaling up 
to the nominal beam energy scale. It is also straightforward to set the copula part 
of the model to the independence copula where the bivariate distribution 
has no dependence.
This parametric model should be relatively easy to adopt in physics event generators and we 
welcome collaboration to facilitate this.
% WHICH IS WHICH ??
%\input{CopulaFit}
%Also discuss copula GoF tests~\cite{Genest}.

%\newpage
\section{Testing New Luminosity Spectrum Modeling}
We generated 20 million events from the fitted copula model and 20 million events 
from a model with no copula (independence copula). The comparison of the center-of-mass energy 
in the luminosity spectrum is shown in Fig.~\ref{fig:copcomp}.
One sees some differences at the 1\% level which we will see has 
some effect on the fitted parameters of the respective luminosity spectrum.
\begin{figure}[!htbp]
\centering
\includegraphics[height=0.45\textheight]{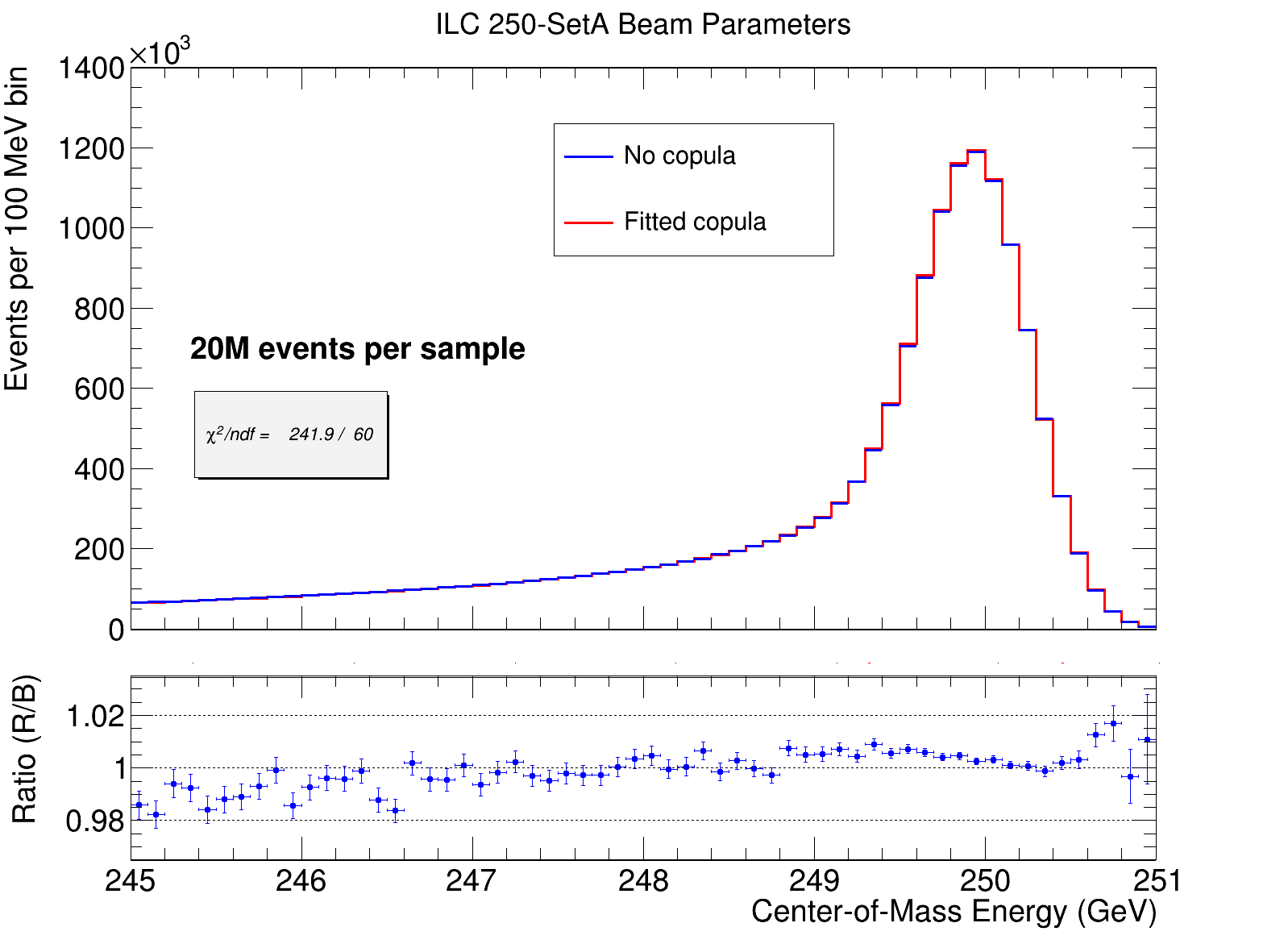}
\caption[]{\small \sl 
Distribution of center-of-mass energy for the two luminosity spectrum models. 
Twenty million events are sampled from each model 
with identically distributed marginal distributions. 
In the no copula model (blue), the electron and positron energy distributions 
in the Body part of the distribution are independent, while for 
the fitted copula model (red) they have a dependence according to the 3-parameter 
copula model fitted to the Guinea-PIG data. The lower panel shows 
the ratio of sampled observations for the fitted copula model to those from 
the no copula model.
}
\label{fig:copcomp}
\end{figure}
Shown in Fig.~\ref{fig:copfits} are fits to the luminosity spectrum for the no 
copula model (top panel) and the fitted copula model (bottom panel) using the 5-parameter 
Gaussian peak and convolved beta tail model described in~\ref{app:GPCBT}. 
One sees satisfactory fits despite the underlying model used 
in the event generation having 17 parameters. 
We believe this indicates that the beam energy spread 
is sufficiently large to smear out the 
fine details needed to model the pre beam energy spread distributions. 
\begin{figure}[!htbp]
\centering
\includegraphics[height=0.45\textheight]{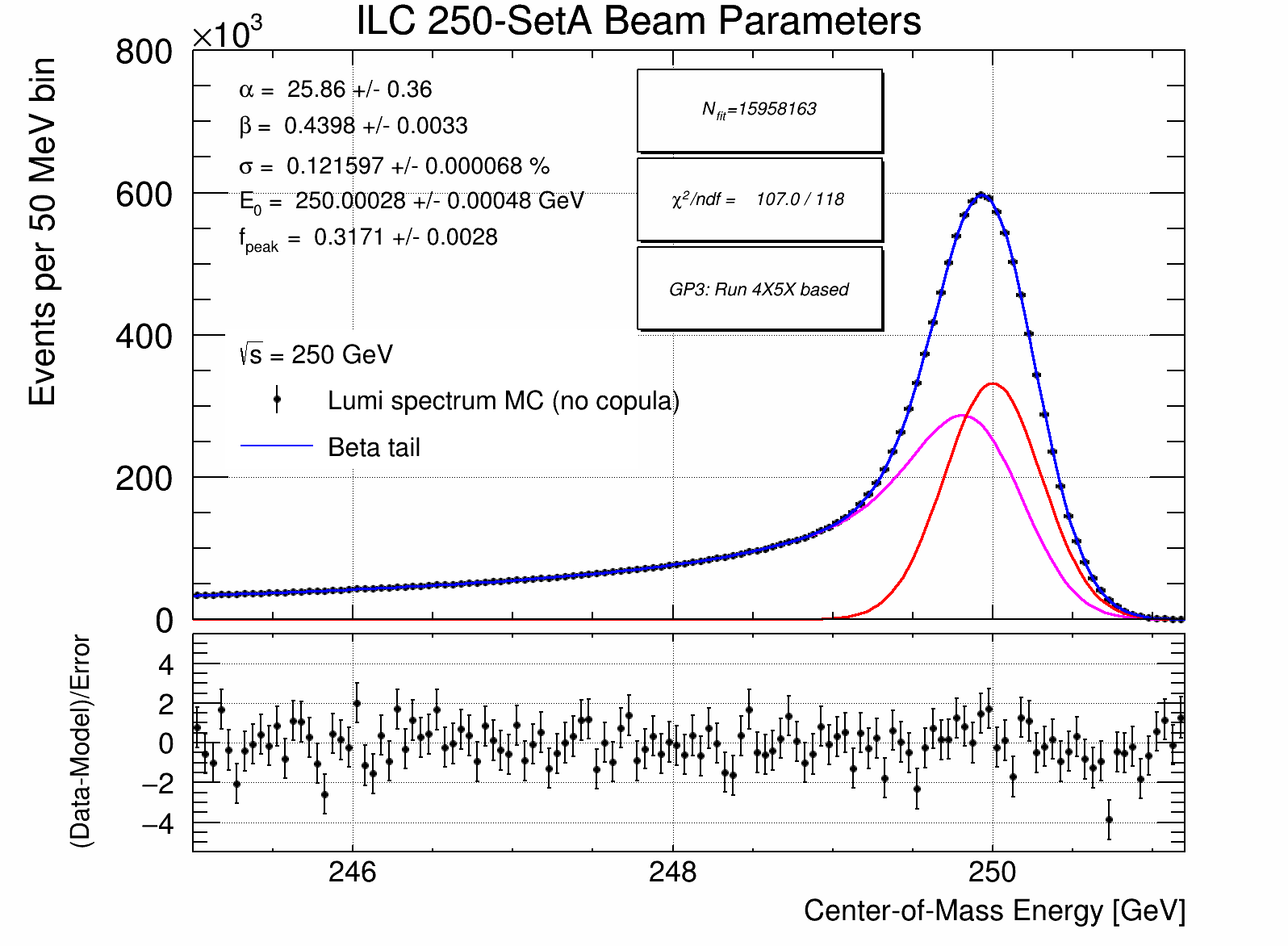}
\includegraphics[height=0.45\textheight]{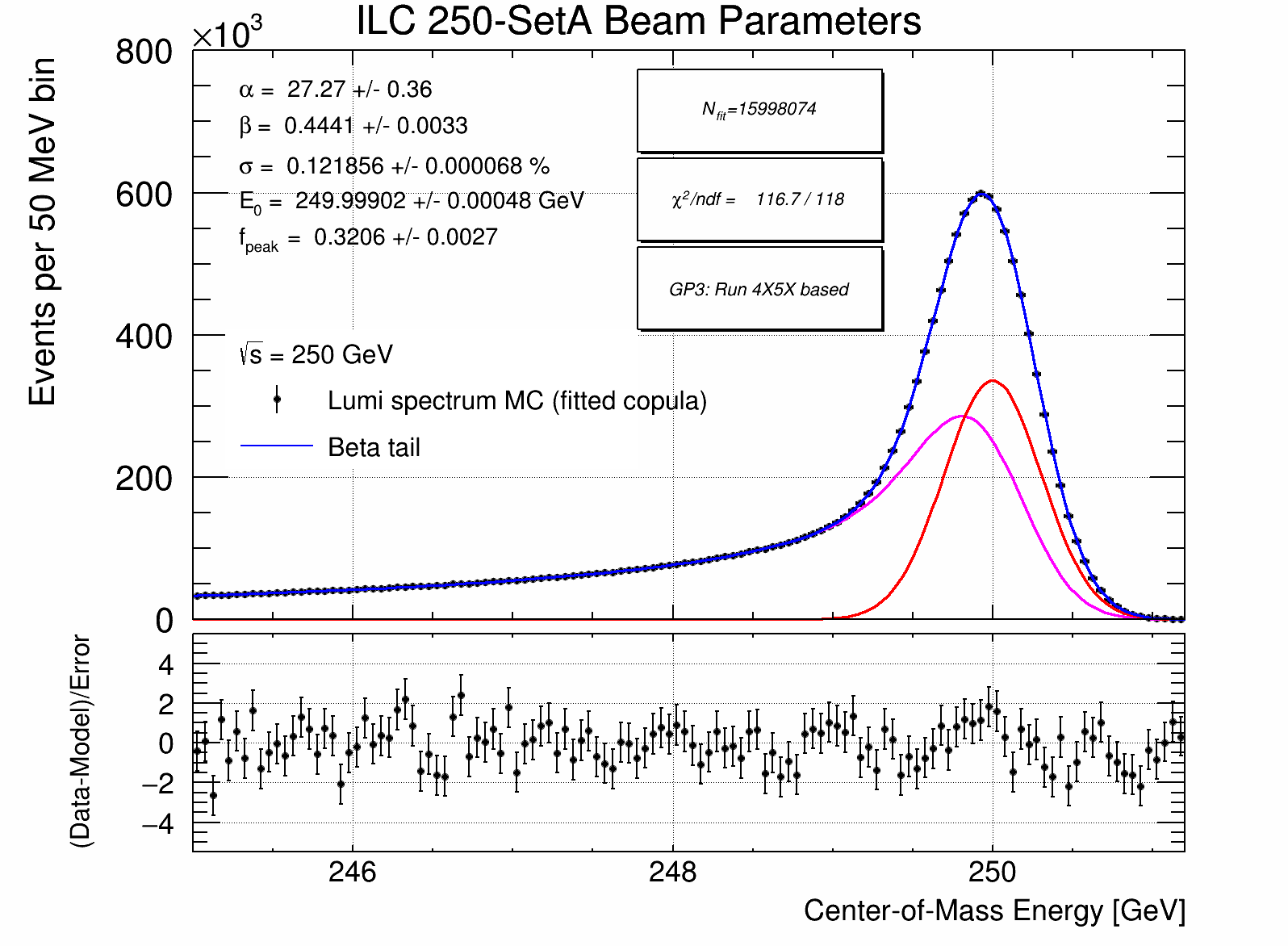}
\caption[]{\small \sl 
Fits to the center-of-mass energy distributions 
for the two luminosity spectrum models of Fig.~\ref{fig:copcomp}.
}
\label{fig:copfits}
\end{figure}
Clearly the fitted parameters from these semi-empirical fits indicate subtle but likely 
significant differences. When each fit is repeated with all but the $E_{0}$ parameter 
fixed to the best fit values for their own data-sets, the fitted energy scale with no 
copula is found to be $5.0\pm0.7$~ppm higher than that for the fitted copula data-set.
Assessed another way, if the non-$E_{0}$ parameters of the fit 
to the fitted copula data-set are fixed and imposed on the fit to the 
no copula data-set, the fitted energy scale with no copula is found to be $1.7\pm0.7$~ppm higher 
than that for the fitted copula data-set, albeit as expected 
with a very poor $\chi^{2}/\nu$ of 375.6/122. A comprehensive evaluation of the energy 
scale systematic from knowledge of the copula parameters was beyond the scope of this work, 
but this initial investigation suggests that if the dependence 
effect can be pinned down to 10\%, 
the consequence for the center-of-mass energy scale determined with dimuons is likely below 1~ppm.
%
%and additionally the fact that some of the finer details for each beam energy distribution 
%will get averaged out in the combined center-of-mass energy quantity.
% One could test this by seeing how well the body and arms center-of-mass energy distributions 
% can be fitted. Do they also need double beta distributions? Not done.
%These have all parameters floating

\section{Detector Momentum Resolution}
The inverse transverse momentum resolution in the ILD 
tracking system for each muon is expected to follow 
approximately a resolution formula given by,
\begin{equation}
\sigma_{1/p_{\text{T}}} = a \oplus b/(p_{\text{T}} \sin{\theta}) \: \: , 
\label{eqn:momres}
\end{equation}
where $\oplus$ 
denotes addition in quadrature.
For a B-field of 3.5~T and 
the full TPC coverage ($37^{\circ} < \theta < 143^{\circ}$), 
the spectrometer 
parameter, $a$, is about $2 \times 10^{-5}$ GeV$^{-1}$, and the 
multiple scattering 
parameter, $b$, is about $1 \times 10^{-3}$.
The modeled dependence is shown in Fig.~\ref{fig:MomRes} 
for different polar angles using fully simulated and 
reconstructed single muon samples using ILD. 
Values for the $a, b$ parameters fitted to momenta of 10~GeV and above 
are reported in Table~\ref{tab:momres}. 
These values for fixed polar angles are then interpolated 
to all polar angles 
in the $7^{\circ} < \theta < 173^{\circ}$ range. 
The maximum observed deviation between 
the simulated data and the parametrization in the fit range is 6\%.
In the future, it would be preferable to update this resolution modeling 
with the latest ILD model, to use more complete sampling of the polar angle range 
especially in the forward region thus minimizing interpolation errors, 
and to investigate more performant momentum resolution possibilities, such as a pixel-based TPC.
\begin{figure}[!htbp]
\centering
\includegraphics[height=0.35\textheight]{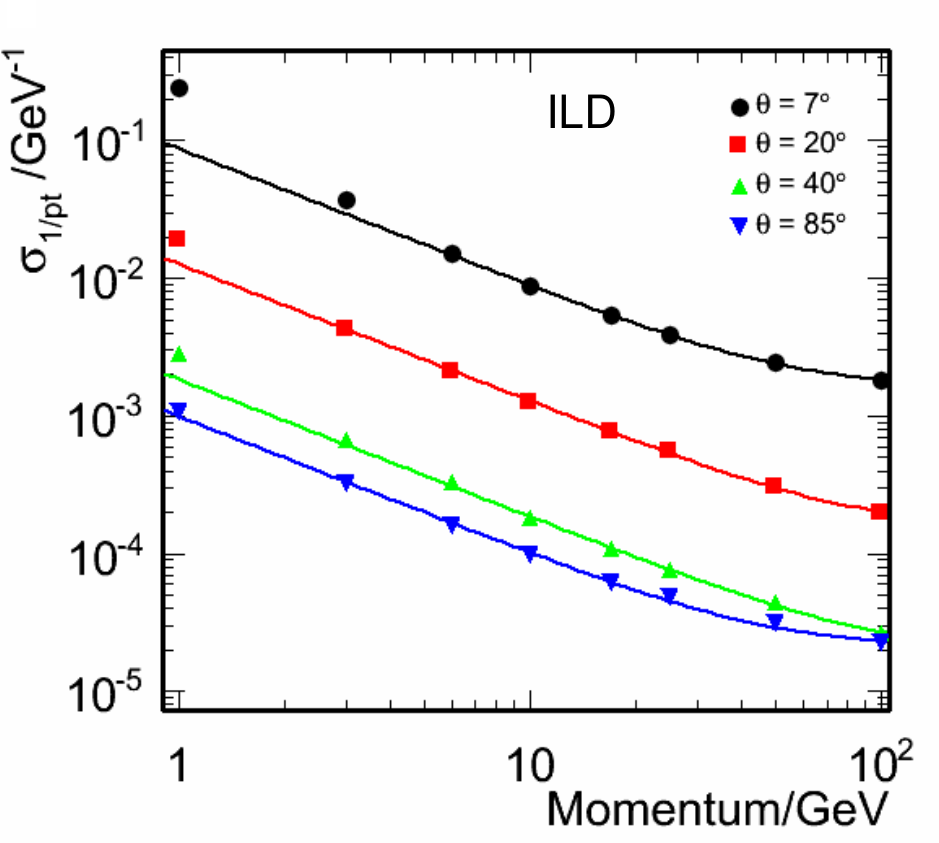}
\caption[]{\small \sl 
Inverse transverse momentum resolution estimated from 
fully simulated and reconstructed single muons at various 
momenta and polar angles for the ILD detector design 
and hit resolutions documented in~\cite{Behnke:2013lya}. 
The superimposed curves are fits over the [10, 100] GeV momentum range 
for the $a, b$ parameters of Eq.~\ref{eqn:momres} at each 
of the displayed polar angles.
}
\label{fig:MomRes}
\end{figure}
\begin{table}[!htbp]
\begin{center}
\begin{tabular}{|r|c|c|}
\hline
$\theta (^{\circ})$ & $a$ ( $10^{-5}$ GeV$^{-1}$) &  $b$ ($10^{-3}$) \\ \hline
85  &  2.09 & 0.978  \\
40  &  1.96 & 0.755  \\ 
20  &  15.4 & 1.48  \\
 7  &  161 &  1.29  \\ 
\hline
\end{tabular}
\end{center}
\caption{
Values for the $a,b$ parameters fitted to the high momentum range 
of Fig.~\ref{fig:MomRes} for four values of polar angle.
}
\label{tab:momres}
\end{table}
%Plot has momentum points at 1,3,6,10,18?,25,50,100

The combined effect of the momentum resolution 
of both muons on the $\sqrtsp$ estimator is non-trivial as the momentum resolution 
depends significantly on the muon polar angle 
especially once tracks no longer benefit from the full radial track length in the TPC.
The resolution is evaluated with a toy Monte Carlo simulation using 
the parametrization of 
the ILD tracking resolution described above.
Figure~\ref{fig:DetRes} shows the intrinsic effect 
of tracking resolution on the reconstructed $\sqrtsp$ estimate 
for two idealized cases as a function of center-of-mass energy, namely,
\begin{enumerate}
    \item The $2 \to 2$ ``full energy'' process ($\ee \to \mumu$) with no additional photons for various scattering angles from $20^{\circ}$ to $90^{\circ}$.
    \item The $\ee \to \mathrm{Z} \gamma \to \mumu \gamma$ ``radiative return'' process where the photon is a collinear ISR photon and the dimuon mass is $\mZ$. The scattering angle in the Z rest frame is set to $90^{\circ}$.
\end{enumerate}
Also included in the figure as a separate component is the intrinsic 
beam energy spread contribution described earlier. 
The fractional resolution on $\sqrtsp$ for full energy events depends markedly on the scattering angle being typically less than 0.2\% for $\sqrt{s} \le 250$~GeV in the barrel.
The radiative return events (light blue curve) have a fractional resolution of 
as good as 0.1\% when both muon tracks are well measured in the barrel. 
This degrades quickly at higher center-of-mass energies as 
the Z is more boosted and the resulting muons from the Z decay are more forward\footnote{For the assumed kinematics and neglecting the muon mass, 
the polar angles of both muons in the lab satisfy  
$|\cos{\theta}| \approx (s-\mZ^2)/(s+\mZ^2)$.}.
%% FIXME add other scattering angles.
%{\bf FIXME Update FCC-ee curve}.
\begin{figure}[!htbp]
\centering
\includegraphics[height=0.40\textheight]{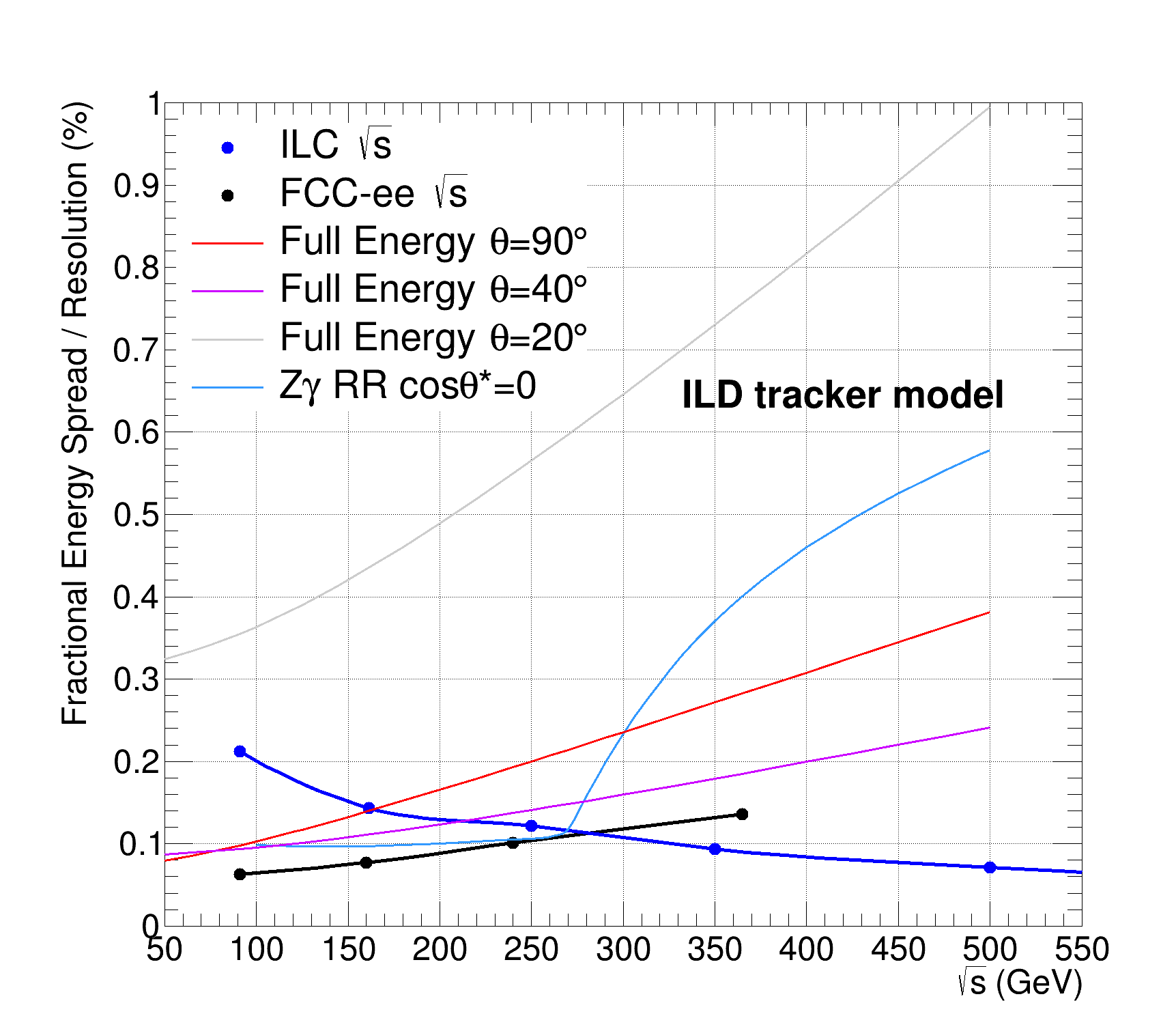}
\caption[]{\small \sl 
Center-of-mass energy dependence of the fractional resolution on $\sqrtsp$ for 
the various configurations described in the text using the ILD tracker model.
}
\label{fig:DetRes}
\end{figure}
\section{Kinematic Acceptance}
Given the strong dependence of the $\sqrtsp$ resolution 
on polar angle, it is important to assess the fraction of 
dimuon events that will pass various acceptance cuts. 
The kinematics of the $\eemmg$ process are characterized 
by the invariant mass of the dimuon system and the 
presence or not of detectable photons. 
The $\sqrtsp$ method is permissive of 
the presence of one photon, but is susceptible to 
errors from events with multiple photons with a
corresponding substantial mass to the photonic system.
One way to elucidate the presence of photons where the photonic system 
has substantial transverse momentum, 
is to examine the acoplanarity angle, 
\[ \phi_{\text{acop}} \equiv \pi - \cos^{-1}\left(  \poneTbf \cdot \ptwoTbf \right) \: \: , \] 
where $\poneTbf, \ptwoTbf$ are unit vectors in the directions of the
transverse momenta of each muon. 
For these detector and physics reasons, it is expected that events with the muons in the barrel acceptance will be better measured 
and that coplanar events with small acoplanarity angles 
will be less sensitive to radiative corrections. 
Of course hard cuts on such quantities will also 
reduce the achievable statistical precision so one approach 
as followed in~\cite{BMadison} is to categorize selected 
events by resolution quality thus avoiding 
unnecessary event loss. 
Figure~\ref{fig:Acceptance} shows an evaluation of 
the acceptance dependence on center-of-mass energy 
for various choices of the acceptance requirements.
One sees that the acceptance varies strongly 
with center-of-mass energy reflecting the substantial fraction 
of events from strong radiative return to much lower 
reduced center-of-mass energies including the Z and below 
where one or more muons are rather forward 
and often undetectable. However, even at $\sqrt{s}=1$~TeV, 
the drop in acceptance for the best measured events compared with $\sqrt{s}=250$~GeV is not that large despite the lack of accepted radiative return to the Z events.
\begin{figure}[!htbp]
\centering
\includegraphics[height=0.35\textheight]{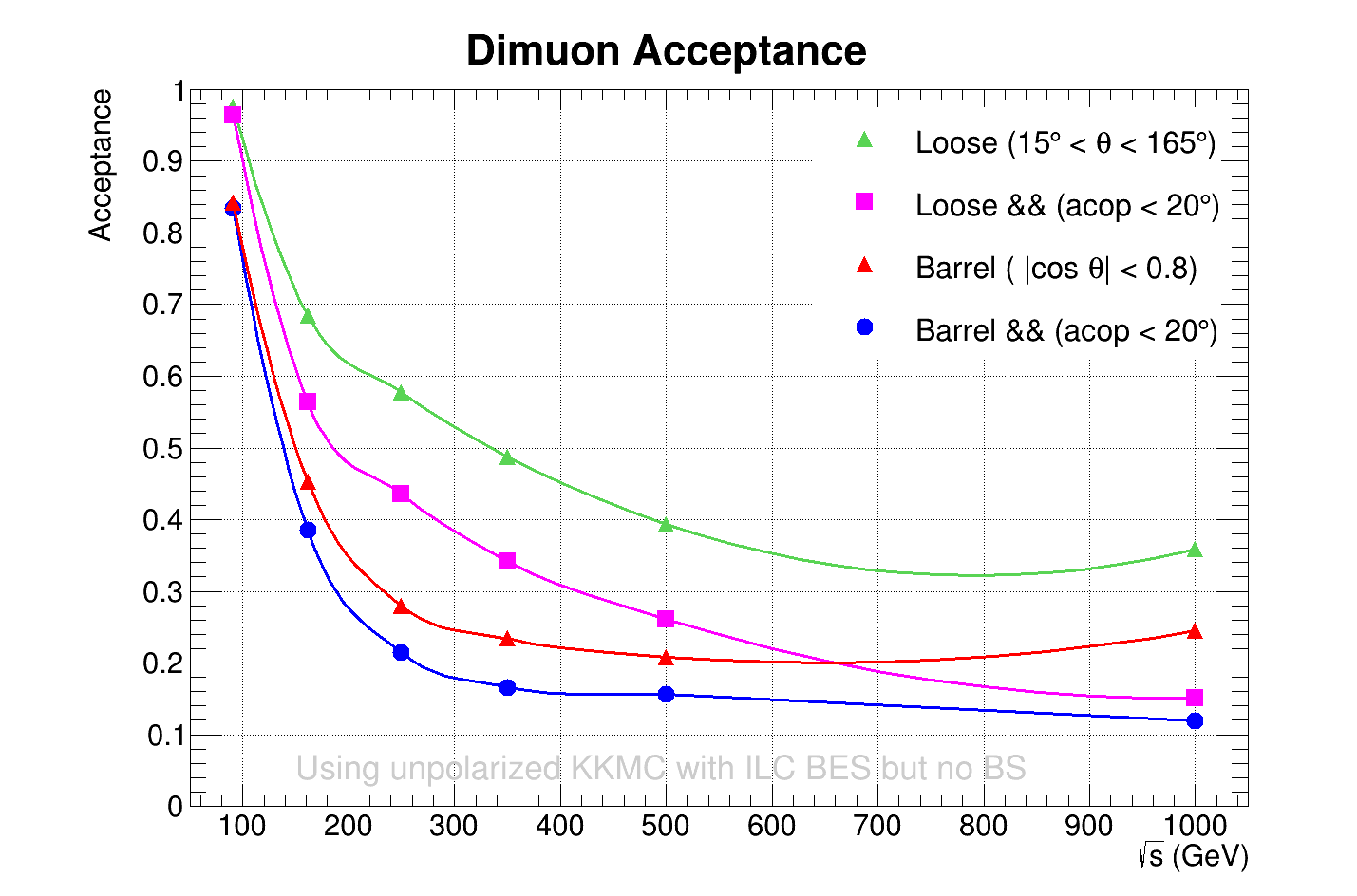}
\caption[]{\small \sl 
Kinematic acceptance dependence on center-of-mass energy 
for $\eemmg$ evaluated using KKMC. Here the 
event generation is unpolarized and includes beam energy 
spread but not beamstrahlung. Shown are acceptance values for 
different polar angle requirements on the muons (loose and barrel).
Also shown is an additional requirement of $\phi_{\text{acop}} < 20^{\circ}$ (shown in purple for loose 
and blue for barrel).
}
\label{fig:Acceptance}
\end{figure}
\section{Physics Limitations}
The physical precision method described in~\cite{KK2f} 
is used to evaluate quantitatively 
how much different levels of higher-order QED effects 
are needed to describe accurately the center-of-mass energy 
related observables.
Weighted events from the KKMCee (4.32) generator~\cite{KKMC} 
were used to re-weight the estimators 
at generator level ($\sqrtsp$, $E^{\text{C}}_{-}$ and $E^{\text{C}}_{+}$) to alternative physics levels. 
%Use $\frac{1}{2} (x^{\prime} - x)$.
The KKMCee event generator uses exclusive 
exponentiation (EEX) and 
coherent exclusive exponentiation (CEEX).
The standard best estimate is ${\cal O} (\alpha^{2}) $ CEEX denoted CEEX2. 
It is also possible to reweight to other models with a 
different, often less complete physics description, 
as follows:
\begin{itemize}
\item ${\cal O} (\alpha) $ CEEX (denoted CEEX1)
\item ${\cal O} (\alpha^{3}) $ EEX (denoted EEX3)
\item ${\cal O} (\alpha^{2}) $ CEEX but no ISR/FSR interference (denoted CNIF2)
\end{itemize}

Table~\ref{tab:physprec} shows the 
$\sqrtsp$ 
physical precision estimates for $\sqrtsp$ in parts per 
million (ppm) for the CEEX1, EEX3, and CNIF2 
variations compared to CEEX2. 
For this comparison, the barrel acceptance 
with a 20$^{\circ}$ acoplanarity requirement (the blue curve of Fig.~\ref{fig:Acceptance}) was used. 
The simulations used 10 million unpolarized weighted events 
per $\sqrt{s}$ and included ILC beam energy spread 
but no beamstrahlung\footnote{
These choices resulted in part 
from experiencing substantial difficulty getting 
an acceptable implementation of polarization effects, beamstrahlung, 
and beam energy spread. This was especially problematic 
for unweighted events. We did fix some obvious errors in 
the beam energy spread implementation. We were not able 
to use the new C$^{++}$ implementation (v5.00.2). 
These issues have 
been reported to the authors.}.
\begin{table}[!htbp]
\begin{center}
\begin{tabular}{|r|c|c|c|}
\hline
$\sqrt{s}$ (GeV) & CEEX1 &  EEX3  & CNIF2  \\ \hline
91          &  -0.9  &  0.4    &   0.02 \\
161         &  -0.2   &  -10   &  -11 \\
250         &  -0.3  &  -11   &  -11 \\
350         &  -0.6  &  -10    &  -10\\
500         &  -0.5  &  -9   &  -9 \\ 
1000        &  -0.6  &  -8    &   -8 \\  \hline
\end{tabular}
% Table 3 acc3 cuts
\end{center}
\caption{
Calculated physical precision in ppm in the $\sqrtsp$ observable vs center-of-mass energy for the CEEX1, EEX3, and CNIF2 variations compared to the CEEX2 setting. 
The mean value for each setting of $(\sqrtsp - \sqrt{s})/\sqrt{s_{\mathrm{nom}}}$ is calculated using a 
$\pm 0.5$\% range. 
The quoted number is half the observed difference 
following the convention in~\cite{KK2f}.
}
\label{tab:physprec}
\end{table}
The recommended comparison is CEEX2 with CEEX1; for 
this case the physical precision for all center-of-mass energies studied is below the ppm level. Studies without the 
acoplanarity cut indicate that the acoplanarity cut 
is needed to achieve this level of physical precision 
for the CEEX2 with CEEX1 comparison especially at high $\sqrt{s}$. For the comparisons with EEX3 and CNIF2 the observed order of magnitude worse physical precision of approximately 10~ppm of Table~\ref{tab:physprec} 
is also found without the acoplanarity cut.
Similar qualitative results to Table~\ref{tab:physprec} were observed 
for the $E^{\mathrm{C}}_{-}$ and $E^{\mathrm{C}}_{+}$ observables but with typically a factor of 3--4 
larger physical precision values calculated for a $\pm 2$\% range.
In conclusion, it appears that the QED effects on the observables studied are already under astonishingly 
good control and for the $\sqrtsp$ observable at a level of one part per 
million (with the acoplanarity angle requirement to suppress higher order effects). 
Therefore, based on these considerations of calculational accuracy, there are excellent prospects 
for including higher-order QED effects with sufficient precision 
in the $\eemmg$ channel.

\section{Bhabhas}
Figure~\ref{fig:Bhabhas} illustrates the statistical 
advantage that is available from Bhabha events compared to dimuon events versus 
center-of-mass energy.
The $\sqrtsp$ method for center-of-mass energy 
relies on precision tracking momentum resolution. 
As an example, 
the barrel Bhabha cross-section 
at $\sqrt{s}=250$~GeV of 24~pb exceeds 
the accepted dimuon barrel cross-section 
of 1.5~pb by a factor of {\it sixteen}. The wide-angle Bhabha cross-sections were evaluated 
with version 1.05 of BHWIDE~\cite{BHWIDE}. 
%In the context of the ILD detector design these barrel events could yield an improvement of four in statistical precision.

Therefore, depending on how well electron tracks 
can be reconstructed compared with muon tracks, 
and their momentum-scale controlled, 
the expected overall sensitivity using $\sqrtsp$ 
can be much better than that from 
the dimuon channel alone such as reported in~\cite{BMadison}.
This is especially true at center-of-mass energies 
above the Z where the $t$-channel enhancement of Bhabha scattering 
at wide angle is most prominent.
In the context of typical detector designs, the still 
much larger cross-section at forward angles needs 
to be tempered by the worsening tracking resolution 
at forward angles beyond the barrel 
region (see for example Fig.~\ref{fig:DetRes} for 
the ILD tracker model). Nevertheless, the statistical precision afforded by Bhabha events will certainly 
be key to improving the monitoring of the time variation of the 
center-of-mass energy scale.

\begin{figure}[!htbp]
\centering
\includegraphics[height=0.35\textheight]{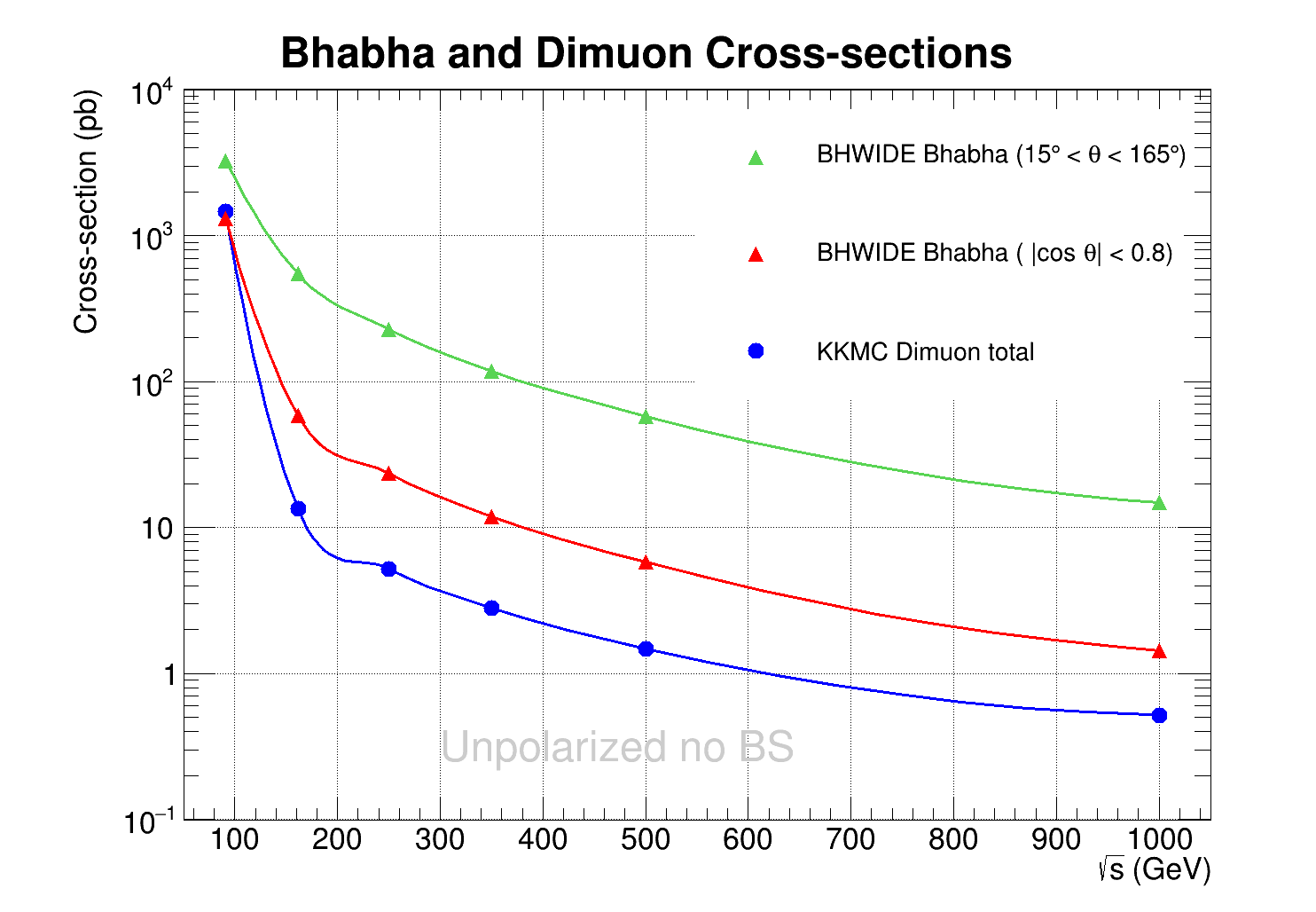}
\caption[]{\small \sl 
Calculated cross-sections in pb vs center-of-mass energy for 
the Bhabha process $\ee \to \ee (\gamma)$ 
for both a barrel acceptance for the detected $\positron$ 
and $\electron$ in red ($|\cos{\theta}| < 0.8$) and a wider acceptance in green ($15^{\circ} < \theta < 165^{\circ}$) using BHWIDE 
and for $\eemmg$ in blue 
using KKMC (no acceptance cuts).
%In all three cases 
The cross-section calculations 
are unpolarized and neglect beamstrahlung and beam energy spread effects.
}
\label{fig:Bhabhas}
\end{figure}

\section{Summary and Outlook}
Developments related to the measurement of 
the center-of-mass energy using dilepton events at future $\ee$ colliders have been described. 
Two new estimators were introduced related to the energy of the colliding electron and positron 
thus permitting more direct study of the 2-d luminosity spectrum marginal distributions. 
Further work is need to fully exploit these.
We developed a parametric model for the 2-d luminosity spectrum that includes modeling 
of the dependence structure using copulas and a simplified treatment of beam energy spread. 
This model should be straightforward to implement in event generators. 

We also surveyed issues 
associated with the $\sqrtsp$ momentum-based center-of-mass energy estimator 
for center-of-mass energies in the 90~GeV to 1~TeV range
including beam energy spread, detector momentum resolution, 
kinematic acceptance, theoretical uncertainties, and the use of 
the Bhabha channel.
Overall we are very encouraged that these techniques can target 1~ppm precision 
on the center-of-mass energy scale thus facilitating 
precision mass measurements. A primary need is control of 
the detector momentum scale at the targeted precision; strategies for this have 
been outlined but will need to be thoroughly assessed.
A full quantitative assessment of the prospects with the $\sqrtsp$ and related methods 
for all $\sqrt{s}$ requires an overall 
consistent estimation procedure including modeling of beamstrahlung, beam energy 
spread, polarization, detector effects, state-of-the-art radiative effects, the inclusion of Bhabhas, 
and an estimator calibration procedure; this was beyond the scope of the current work.
\label{sec:concl}
\section*{Acknowledgments}
I acknowledge fruitful discussions with Andr\'e Sailer and 
Daniel Schulte on the use of Guinea-PIG and with Brendon Madison on many 
aspects of this endeavor.
This work is partially supported by the US National Science Foundation 
under award NSF 2013007. 
This work benefited from use of the HPC facilities operated 
by the Center for Research Computing at the University of Kansas 
including those funded by NSF MRI award 2117449. 
I would also like to thank the LCC generator working group and 
the ILD software working group for providing the 
simulation and reconstruction tools and producing 
some of the Monte Carlo samples used in this study.
This work has also benefited from computing services provided by 
the ILC Virtual Organization, supported by the national resource 
providers of the EGI Federation and the Open Science GRID.
%This version is a copy of mybib.tex 
%as of version distributed to Alberto.

\appendix
\section{Fit models}
\label{app:fits}
We have used a new fit model to 
characterize the expected sensitivity to 
the center-of-mass energy scale parameter and to model 
the shape of related distributions. 
This new model 
is based on convolving a CIRCE-like function 
with a Gaussian response function to model beam energy spread or detector resolution.
In our earlier work~\cite{BMadison} we had used two fit models: the well known Crystal Ball function, and the convolution 
of a Gaussian response function with a mixture of 
a two-component exponential for the low energy tail and a delta function for the peak. 
The latter more flexible fit model was chosen partly for convenience since it 
is analytically integrable. Details of the latest model and its 
implementation using numerical integration are described below.
For completeness we also describe the fit model used for the pre beam energy spread distributions.

\subsection{Gaussian peak and convolved beta tail}
\label{app:GPCBT}
%\subsection*{Gaussian Convolved CIRCE-like Model}
Previous work on fitting the electron and positron 
energy distributions absent beam energy spread effects 
had been done using a mixture of a beta distribution for the beamstrahlung tail and a delta function for 
the undisturbed peak in CIRCE \cite{Ohl:1996fi}. 
Here we apply similar methodology to either the 
center-of-mass energy or single beam energy distributions but 
taking into account a Gaussian response function using convolution.
We define the scaled true energy, $x = E/E_{0}$, where $E$ 
is the true energy, and $E_{0}$ is the energy scale 
parameter.
The resulting probability density function of the fit model at convolved 
scaled energy, $\xp = \Ep/E_{0}$, has 5 parameters, 
and is given by  
\begin{multline}
    p(\xp; \alpha, \beta, \sigma, E_{0}, \fpeak) = \int_{0}^{1}
     [\fpeak \delta(x - 1) + \frac{(1-\fpeak)}{B(\alpha, \beta)} 
     x^{\alpha-1}(1-x)^{\beta-1}
     ] \; G(\xp-x; \; \sigma)  dx \; \text{,}  
    \label{eqn:conv}
\end{multline}
where $\Ep$ is the convolved energy, $\alpha$ and $\beta$ are the beta 
distribution parameters, $\fpeak$ is the fraction associated with the delta function component located at $x=1$, 
$\sigma$ is the standard deviation of the Gaussian response function as a fraction, and 
the normalizing coefficient is
\begin{equation}
B(\alpha, \beta) = \frac{\Gamma(\alpha) \Gamma(\beta)} { \Gamma (\alpha + \beta) } \: ,
\end{equation}
where $\Gamma$ is the Gamma function.
The Gaussian response function is simply,
\begin{equation}
    G(\xp - x ; \; \sigma) = \frac{1}{\sqrt{2 \pi \sigma^2}} \exp\left( - \frac{(\xp - x)^{2}} {2 \sigma^2} \right) \; \text{,}
\end{equation}
which smears deviations from the true scaled energy, $x$, of size, $\xp - x$, 
with resolution, $\sigma$. 

As in CIRCE, a variable transformation is used 
to take care of the integrable 
singularity of the beta distribution at $x=1$.
%We cannot simply restrict the window of integration 
%as the distribution in $\sqrt{s}$ goes above $\sqrt{s_{nom.}}$ due to beam energy spread. 
%We have implemented this using RooFit~\cite{RooFit}. 
%and fit using MINUIT's binned log likelihood minimizer \cite{RooFit} \cite{MINUIT}. 
Multiple numerical integration methods for the convolution were tested particularly 
with a view to the stability and speed of the fit procedure.
%For this work Boole's rule for $N=180$ subintervals per bin was used.
For this work 30-point Gauss-Legendre integration 
over a $\pm 6 \sigma$ integration window was used 
for the probability density evaluation and the fit was 
implemented using RooFit~\cite{RooFit}.

This 5-parameter fit function manages to describe the simulated Guinea-PIG data, which includes beam energy spread effects, 
much better or with fewer parameters than the two prior 
models. Therefore we have 
adopted it as the preferred empirical fit model for center-of-mass energy and beam energy distributions. 

\subsection{Double beta tail}
\label{app:DBT}
For the fits to the Body and Arms regions of the pre beam 
energy spread marginal distributions we fit a 5-parameter 
two-component model with two beta functions to 
each $x$ distribution as follows
\begin{multline}
    p(x; \alpha_1, \beta_1, \alpha_2, \beta_2, f_{1}) = 
   \frac{f_1}{B(\alpha_1, \beta_1)} x^{\alpha_1-1}(1-x)^{\beta_1-1} 
   + \frac{(1-f_1)}{B(\alpha_2, \beta_2)} x^{\alpha_2-1}(1-x)^{\beta_2-1}  \: \: \text{.}  
    \label{eqn:conv2}
\end{multline}

\end{document}